\documentclass[12pt]{article}
\usepackage[dvips]{graphicx}
\usepackage{pdproc}
\usepackage{wrapfig}

\newcommand{\mgl}{m_{\tilde{g}}}
\newcommand{\msb}{m_{\tilde{b}}}

  \makeatletter 
  \def\@cite#1{[#1]} 
  \makeatother    
\begin{document}

\renewcommand{\thefootnote}{\alph{footnote}}

\title{
Supersymmetry at the LHC and Beyond\\
--Ultimate Targets---
}

\author{ Mihoko M. Nojiri}

\address{   
YITP, Kyoto University \\
Sakyo, Kyoto, 606-8502 Japan
\\ {\rm E-mail: nojiri@yukawa.kyoto-u.ac.jp}}

\abstract{
At the LHC, superpartners with 
the masses lighter than a few TeV may be found, and  the masses 
of the supersymmetric (SUSY) particles and their interactions will  
be studied. 
The information will be the base to consider  
the SUSY breaking mechanism and cosmology 
related to  the SUSY dark matter. We review 
the recent studies that aims for  the ``precise SUSY study''
 at the LHC.
}

\normalsize\baselineskip=15pt

\section{Introduction}

The LHC experiment, scheduled to start from 2007, is the best place to
study the supersymmetry in coming 10 years. The LHC is a pp collider at
$\sqrt{s}=14$~TeV. It is able to find $\tilde{q}$ and $\tilde{g}$ 
in the minimal supersymmetric standard model(MSSM) up
to 2--3~TeV. Not only for that, it is also
the place to do the precise measurement of the nature of the SUSY
particles.  If the sparticle masses are as light as 1~TeV, the
production cross section is as large as 3~pb, therefore O(10000) events
/year will be produced for the detailed  study.  It is our 
responsibility
to  extract physics  information, as much as possible, from the 
LHC experiment for the future developments of the experimental 
and theoretical particles physics. 

Different SUSY breaking models leave special imprints to the mass
spectrums, therefore measurement of the masses of the SUSY particles
is a important target for the collider physics.
After finding superpartners, we can select  pure samples of
the SUSY events free from standard model (SM) backgrounds.  By looking into the
distributions of the sparticle decay products, we can determine the
mass of the sparticles, which may provide us the understanding of the
SUSY breaking mechanism, namely, physics at the very high scale.

The measurements also have impacts to the other physics.  The mass
spectrum determination constrains the SUSY contributions to the flavor
violating processes, such as b rare decays, $\mu\to e\gamma$, and
muon  anomalous magnetic moment.  The thermal relic density of the
SUSY dark matter also can be calculated if the MSSM parameters 
are known.

One may  even  study  some of the SUSY
relations. In the MSSM, the chiral structure of the coupling of the
superpartners are restricted. For example the bino and wino couplings
are the (s)fermion chirality conserved form
$f_{L(R)}\tilde{f}_{L(R)}\tilde{B}$ or
$f_{L}\tilde{f}_{L}\tilde{W}$. The sparticle left-right mixing are
induced by the $F$ term as $(m^2)_{\tilde{l}}^{LR}\propto
m_l\mu\tan\beta$. We will discuss how to prove such relations at the
LHC for explicit examples.

Before going into the detail,  
let us first summarize   basic features 
of the collider signatures of the SUSY  particles at the LHC. 
In the models where SUSY is broken at 
high energy scale, the squark and gluino masses 
are much higher than that of the weak interacting SUSY 
particles.  Therefore, at least one jet originated 
from a squark or a  gluino  
cascade decay has large $p_T$. The decay products 
include  weakly interacting SUSY particles, which further decay 
into the leptons or jets. Especially, at the end of  each 
sparticle cascade decay, the 
lightest SUSY particle (LSP) appears, which may be stable due to the 
R parity conservation. The LSP cannot be detected in the detector,
therefore SUSY events have large missing momentum. By applying the 
cuts on the missing $P_T$ and the total transverse energy $M_{\rm eff}$, 
the SM backgrounds can be reduced to the negligible level.

One of the relatively clean channels is the 2 body 
cascades or 3 body decays of the heavier neutralinos
$\tilde{\chi}^0_i$
 which  have  been obtained lots of attentions 
in the past physics performance studies\cite{atltdr}, 
\begin{equation}
\tilde{\chi}^0_i\rightarrow (\tilde{l}l) \rightarrow ll\tilde{\chi}^0_1.
\label{chidecay}
\end{equation}
The cascade decay produce opposite sign same flavor  leptons(OSSF). 
The accidental 
leptons can be estimated from the $e\mu$ distribution 
(opposite sign opposite flavor leptons---OSOF), 
leaving very clean samples containing $\tilde{\chi}^0_i$.

Another important SUSY channels are those  containing $b$ quarks.
In many SUSY breaking models,  $\tilde{b}$ and $\tilde{t}$ could 
be much lighter than other squarks because 
of the negative running of the squark mass by the Yukawa coupling, 
in addtion to the left-right mixing. The decays 
$\tilde{g}\rightarrow
\tilde{b}, \tilde{t}$ are to open in the wide parameter space. 
The LHC detectors have a good tagging efficiency for  $b$ jets $\epsilon_b\sim
0.5$, and the cascade decays,
\begin{eqnarray}
\tilde{g}&\rightarrow& b\tilde{b}\ \ 
\left\{\begin{array}{l}
\tilde{b}\rightarrow \tilde{\chi}^0_ib\, \cr
\tilde{b}\rightarrow \tilde{\chi}^+_it\, 
\end{array}
\right.
\cr
\tilde{g}&\rightarrow& t \tilde{t} \ \ 
\left\{\begin{array}{l}
\tilde{t}\rightarrow \tilde{\chi}^0_i t\, 
\cr
\tilde{t}\rightarrow \tilde{\chi}^+_i b\, 
\end{array}
\right.
\end{eqnarray} 
maybe studied at the LHC by tagging two $b$ jets. 
In the next section, we will discuss 
these processes as well. 

\section{Sparticle Mass Measurements at the LHC}

If the LSP is  stable,
the SUSY events have large missing transverse
momentum.  This is important signature to discriminate it from the SM 
backgrounds.  On this other hand, 
one has to determine the  parent SUSY particle masses
from the momentum of the visible particles only.  This is easy for the
case of $e^+e^-$ collider experiments.
The energy of collisions are
known, and the decay kinematics can be solved by the constraint. 
But  it is not simple task at the LHC,
where a proton is a composite particle and collision
energies of the partons are  not fixed.

The techniques to do the  jobs at the LHC have been  developed in 
the past 10 years.
We quickly review the previous works and also explain new ideas.

\subsection{The case with the long lived NLSP}
In the models with the gravitino LSP, the NLSP is 
long lived and could be either charged or 
neutral. The NLSP candidates are  sleptons (especially 
the lighter stau $\tilde{\tau}_1$), 
the lightest neutralino $\tilde{\chi}^0_1$, or sneutrinos $\tilde{\nu}$. 
The NLSP could decay into its SM partner and a gravitino
\begin{eqnarray}
\tilde{l}&\to& l \tilde{\psi}_{3/2},
\cr
\tilde{\chi}^0_1&\to& \gamma \tilde{\psi}_{3/2}.
\end{eqnarray}
 One can extract the mass of SUSY particles by measuring the endpoints 
of the invariant mass distributions of the selected particles 
involving  the photon or lepton. 
(The endpoint method is discussed in the next
section).  For example, from the endpoint of the $m_{\gamma l}$,
$m_{\gamma ll}$ and $m_{ll}$ ..  of the cascade decay
$\tilde{\chi}^0_2\rightarrow \tilde{l}l\rightarrow \tilde{\chi}^0_1 ll
\rightarrow \tilde{\psi}_{3/2} ll\gamma$, 
one can calculate  $\tilde{\chi}^0_{1(2)}$ and $\tilde{l}$ 
mass model independently with high precision of 
$O(1)$~GeV. The structure of the messenger sector will be determined 
very precisely from the mass measurements.

The life time of the NLSP depends on the scale of the hidden sector
SUSY breaking $F$ which is also related to the gravitino mass
$m_{3/2}=F/({\sqrt{3}M_{\rm pl}})$.  We may learn the structure of the
hidden sector from the measurement of the life time.  For example, the
life time is given for the decay $\tilde{l}\to l \tilde{\psi}_{3/2}$
as \cite{Giudice:1998bp}
\begin{equation}
\tau_{\rm NLSP}=\left(\frac{100{\rm GeV}}{m_{\rm NLSP}}\right)^5
\left(\frac{\sqrt{F}}{100{\rm TeV}}\right)^4
3\times 10^{-13}{\rm sec}.
\end{equation}

The gravitino could be a dark matter. It can be produced either
thermally or from the NLSP decays.  However if the life time of the
NLSP is longer than 1 sec, the decay products may affect the BBN and
effect depends on the life time strongly. The cosmology related to the
gravitino is discussed in this workshop in detail\cite{Feng,Hama}.

The signature of the charged NLSP(CNLSP) would be very spectacular. 
It 
is detected as a highly ionizing particles as
it is non-relativistic. The time of the flight(TOF) information at the muon
systems and momentum measurement at the inner detector allows precise
measurement of the NLSP mass\cite{Hinchliffe:1998ys}.  All neutralino 
masses may be  determined within the error of  O(1)~GeV
by combining the NLSP momentum with that of the other 
visible particles. 

The supergravity  interaction may also  be studied in detail
for this case.  The charged NLSP may travel through the detector with
finite vertex until it decays finally.  The life time can be
calculated from the distribution of the flight distance and the decay
time of the NLSP. It should be possible to do such measurements as far
as they decays inside the detector, although there are no systematic
experimental study on the expected precision of the lifetime at the
LHC.  Even if $c\tau_{\rm NLSP}$ is longer than the detector scale, the stable
charged particle may be stopped at the massive stopper placed nearby
the main detector\cite{Feng}.  Recent study shows that one can
determine the life time as long as O(100) years at LHC,  if a massive
stopper with mass around 1 kton can be placed near the LHC
detectors\cite{Hamaguchi:2004df,Feng:2004yi}.

For the neutral long lived NLSP, the NLSP momentum cannot be measured
directly. If the life time is short enough, it decays into a photon
and a gravitino in the detector, and the photon will be detected at
the ECAL.  The reconstruction of the NLSP momentum, a decay position
and a decay time of the NLSP is possible event by event by using the
time information and photon momentum measurement at
ECAL\cite{Kawagoe:2003jv}.  This is because non-relativistic
neutralinos fly in the detector then decay, therefore the arrival time
depends on both the NLSP momentum and decay time.  The resolution of
the decay time and position depends on the time resolution of the
ECAL. The ECAL time resolution is about
O(0.1)~nsec in a beam test, 
which corresponds to O(3cm) resolution on the NLSP
vertex.  Assuming the time resolution, the mass error obtained from
the reconstructed NLSP momentum is estimated  as low as O(1)~GeV
at a model point\cite{Kawagoe:2003jv}, for
the process
\begin{equation}
\tilde{l}\to \tilde{\chi}^0_1 l \to\tilde{\psi}_{3/2}\gamma l.
\end{equation}

In the limit where the NLSP life time is too long to decay in the
detector, the signature becomes very close to the case where the LSP
is the lightest neutralino.  This will be discussed in the next
subsections.

\subsection{The case with the lightest neutralino LSP at the LHC}
When the lightest neutralino is  stable,  it is not possible  
to measure the momentum directly. The study of the mass spectrum 
is still possible by looking into the decay distribution of the 
visible particles from sparticle decays. 

The past physics studies are mostly done with  the endpoint method. 
Namely,one measures 
endpoints of the invariant mass distribution of 
the selected sample which are dominated by the events 
from a decay cascades. The positions of the endpoints 
depend on the sparticle masses. 
For example, the endpoint of the $m(l^+l^-)$  distribution 
for the cascade decay  given in Eq.~(\ref{chidecay}) is 
\begin{eqnarray}
m_{ll}({\rm 3\ body})&=& m_{\tilde{\chi}^0_2} -m_{\tilde{\chi}^0_1}, \ \ \
\cr
\cr
m_{ll}{\rm (2\ body)}&=& \sqrt{\frac{
(m^2_{\tilde{\chi}^0_2}-m^2_{\tilde{l}})
(m^2_{\tilde{l}}-m^2_{\tilde{\chi}^0_1} )}
{m^2_{\tilde{l}}}},
\end{eqnarray}
respectively. One can then select a jet from a squark decay 
$\tilde{q}\rightarrow 
q\tilde{\chi}^0_2$ by taking one of the two highest $p_T$ jets. 
The endpoint of the 
invariant mass of the jet and leptons 
provide the constraints among  
$m_{\tilde{q}}$ and the lighter SUSY particle
masses involved in the cascade decay. 

For the point SPS1a which is defined in the mSUGRA scenario by the
parameters $\tan\beta=10$, $m=100$~GeV, $M=250$~GeV, $A_0=-100$~GeV,
and $\mu>0$
\cite{Allanach:2002nj},
the error of sparticle masses at the LHC for the integrated luminosity 
$\int {\cal L} dt=300$~fb$^{-1}$
are estimated in \cite{gjelsten}, and the results are summarized in 
Table~\ref{errorsps}.
\begin{table}[tbh]
\begin{tabular}{|c|c|c|c|l|}
\hline
sparticle & mass(GeV) &  error (GeV)&with $m_{\tilde{\chi}^0_1}$
input & comment\cr
\hline
$\tilde{g}$       &595&8.0&6.4&$bbll$ mode\cr
$\tilde{q}_R$     &520&11.8&10.9&use $M_{T2}$\cr
$\tilde{q}_L$     &540&8.7&4.9&$jll$ mode\cr
\hline
$\tilde{b}_1$     &492&7.5&5.7&$bbll$, need collect $b$ jet calibrations \cr
$\tilde{b}_2$     &525&7.9&6.3& to separate two $\tilde{b}$'s \cr
\hline
$\tilde{\chi}^0_4$&378&5.1&2.25&high $ll$ edge\cr
$\tilde{\chi}^0_2$&177&4.7&0.24&$jll$ mode\cr
$\tilde{l}_R$     & 143&4.8&0.34&$jll$ \cr
$\tilde{\chi}^0_1$& 96&4.8&N.A.&$jll$\cr
\hline
\end{tabular}
\caption{Mass of the sparticles at SPS1a, and expected sensitivity 
at the LHC, and that  with the expected input from the ILC.}
\label{errorsps}
\end{table}

The cross section at SPS1a is large ($\sigma\sim 58$pb) and 
$O(10^6)$ sparticle will be produced at the high luminosity 
option of the LHC at this point. 
For weakly interacting particles, the errors are dominated by 
the uncertainty of the absolute mass scale, 
while the mass differences of the sparticles 
are more strongly constrained. This is general 
feature of the determination of the sparticle 
masses when the LSP momentum cannot be measured directly.
The mass of the lightest neutralino may be measured more precisely 
at the international linear collider (ILC), then  
error of the sparticle masses would be reduced significantly
as can be see the 4th column of the Table~\ref{errorsps}.
Finally the errors for the squark and gluino masses 
are dominated by the 1\% jet scaling error.  

It is notable that the LHC can access to the mass of the the heaviest
neutralino $\tilde{\chi}^0_4$ through its decay into sleptons. For
this model $m_{\tilde{\chi}^0_4}=378$~GeV, which 
 cannot be accessible  at
$\sqrt{s}=500$~GeV $e^+e^-$ colliders. 
 The measured mass difference $m_{\tilde{\chi}^0_4}-m_{\tilde{\chi}^0_1}$
directly constrains $\mu$ parameter.  The errors for the MSSM
parameters would be significantly reduced 
when the measurements at the LHC and the ILC
can be combined\cite{Desch:2003vw}.

We also can see the implication to the cosmology.  The mass of the
slepton and the lightest neutralino can be determined within O(5)\%
level. This means one can calculate the neutralino pair annihilation
rates into the first and second generation leptons  
at the time of decoupling with in the error of 10\%. The 
pair annihilation rates into $\tilde{\tau}$ is also important. The
$\tilde{\tau}$ mass also can be measured at the LHC, but
more study is needed to obtain the error, as the channel 
involving $\tau$ leptons suffers backgrounds from QCD jets. 

\subsection{Mass relation method}
Although the mass determination through  the endpoint method is
successful, there are  
problems which may limit the application. The problems
may be summarize as follows;
\begin{itemize}
\item The LSP momentum cannot be reconstructed except 
for a few very special part   of the decay phase space. 
\item 
Only the 
events near the endpoints  of the decay distribution 
are used for the mass fit. The large 
statistics are required to see the endpoints 
while  the events away from  the endpoints 
also contain the independent information of the 
masses.

\item 
The selected SUSY events may contain events from several cascade
chains.  For example, three or four channels involving $\tilde{b}_i$
or $\tilde{t}_i$ contribute simultaneously for $bbll$ final state.
The measured $m(bbll)$, $m(bll)$ and $m(bl)$ endpoints are the
weighted average of the endpoints of these channels. This introduces
additional systematical uncertainty for the mass determinations.
\end{itemize}

The above problems  
may be removed by using the ``mass relation method''
\cite{Nojiri:2003tu,Kawagoe:2004rz} 
when a sparticle cascade decay chain is sufficiently 
long. 
In the mass relation method, one  uses the on-shell condition of the sparticle 
masses to solve the kinematics of the sparticle decay product exactly, 
and reconstruct the masses of the SUSY particles as the peaks  of  
certain distributions.  
To describe the method, let us take the process
\begin{equation}
\tilde{g}\rightarrow \tilde{b}b_2\rightarrow \tilde{\chi}^0_2 b_1b_2
\rightarrow \tilde{\ell}b_1b_2\ell_2 \rightarrow \tilde{\chi}^0_1b_1b_2\ell_1\ell_2,
\label{bbll} 
\end{equation}
where $b_1$, $b_2$, $l_1$ and $l_2$ denote different b quarks and
leptons respectively.  Both of the sbottom states $\tilde b_1$ and
$\tilde b_2$ yield the decay chain of Eq.~(\ref{bbll}).

\begin{table}
\begin{center}
\begin{tabular}{|c|c|c|c|c|c|}
\hline
 $\tan\beta$ &
$m_{\tilde{g}}$ & $m_{\tilde{b}_{1(2)}}$ & $m_{\tilde{\chi}^0_2}$ & 
$m_{\tilde{\ell}_R}$ 
&$m_{\tilde{\chi}^0_1}$ \cr
\hline
10 & 595.2& 491.9 (524.6)& 176.8&  143.0& 96.0\cr
\hline
15 & 595.2& 485.3 (526.9)& 177.9&  143.0& 96.5\cr
\hline
20 & 595.2& 478.7 (531.2)& 178.5&  143.1& 96.7\cr
\hline
\end{tabular}
\caption{Masses of some sparticles at  SPS1a and the 
variants studied in Section 3, in GeV.}
\end{center}
\label{table1}
\end{table}

Five sparticles are involved in the cascade decay Eq.~(\ref{bbll}),
therefore one can write five mass shell conditions among the leptons
and quarks in the final decay products.
\begin{eqnarray}
m^2_{\tilde{\chi}^0_1}&=& p^2_{\tilde{\chi}^0_1},\cr
m^2_{\tilde{\ell}}&=& (p_{\tilde{\chi}^0_1}+ p_{\ell_1})^2,\cr
m^2_{\tilde{\chi}^0_2}&=& (p_{\tilde{\chi}^0_1}+
p_{\ell_1}+p_{\ell_2})^2,\cr
m^2_{\tilde{b}}&=& (p_{\tilde{\chi}^0_1}+
p_{\ell_1}+p_{\ell_2}+p_{b_1})^2,\cr
m^2_{\tilde{g}}&=& (p_{\tilde{\chi}^0_1}+ 
p_{\ell_1}+p_{\ell_2}+p_{b_1}+p_{b_2})^2.
\label{gluino}
\end{eqnarray}

For a $bb\ell\ell$ event, the equations contain the 4 unknown degrees
of freedom of the $\tilde{\chi}^0_1$ momentum.  Each event therefore
describes a 4-dimensional hyper-surface in a 5-dimensional mass
parameter space, and the hyper-surface differs event by event. From
the purely mathematical point of view 5 events would be enough to
determine a discrete set of solutions for the masses of the involved
sparticles.   

From now on, we will develop the argument by assuming that the masses
of $\tilde{\chi}^0_2$, $\tilde{\ell}$, and $\tilde{\chi}^0_1$ are
known for simplicity.  This is a reasonable assumption at the LHC,
where it has been shown that a detailed study of the lepton-lepton
system from the $\tilde{\chi}^0_2$ decay can be used to precisely
constrain these masses. In addition to that, if the international
linear collider (ILC) would  be build, the mass of lighter sparticles
will be measured with less than 1\% accuracy there.  If we assume the
mass of the lighter sparticles is known, each event corresponds to a
different curve in the $(\mgl,\msb)$ plane, which is expressed as
following equation;
\begin{equation}
 f(\mgl,\msb, p_{b1}, p_{b2}, p_{l1}, p_{l2})
=Q_{11} \mgl^4+2Q_{12} \mgl^2 \msb^2+ Q_{22}\msb^4
+2Q_{1} \mgl^2+ 2Q_{2}\msb^2 + Q=0,
\label{maineq}
\end{equation}
where $Q'$s are the functions  of the lighter sparticle
masses  and 
momenta of the $b$ quarks and leptons. 
Two events are enough to solve the gluino and sbottom masses
altogether up to the maximally  four solutions.
\footnote{In addition, there are two possibility 
to assign the two leptons to  $l_1$ and $l_2$.
In the plot of this paper, we fix the assignment 
by fixing the assignment so that the lepton with the higher 
$p_T$ is $l_1$.}

\begin{wrapfigure}[30]{l}{0.4\textwidth}
\centerline{
\includegraphics[width=6cm]{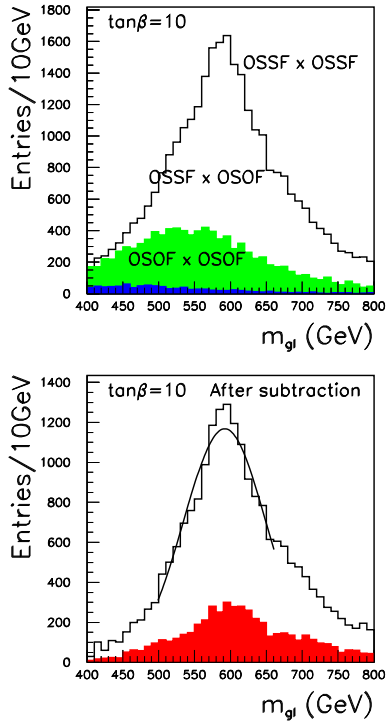}
}
\caption{The gluino mass distributions for
SPS1a  with the event pair analysis.
The open, green, and blue histograms
in the top figures are for OSSF$\times$OSSF,
OSSF$\times$OSOF, and OSOF$\times$OSOF event pairs,
respectively.
The open histograms in the bottom figures
shows the mass distributions after background subtraction.
The contributions of $\tilde{b}_2$ are shown by red histograms.
}
\label{mglplot}
\end{wrapfigure}

We call this technique ``mass relation method'', because here one
uses the fact that the sparticle masses are common for all events which go
though the same cascade decay chain.  Note that the events need not to
be close to the endpoint of the decay distribution, but they are still
relevant to the mass determination. This means that one can use the
mass relation method even if the number of signal events is small.

In the following, we start from the selected
$bb\ell\ell$ events at SPS1a.  
The decay distribution is studied by generating
events by using HERWIG\cite{Corcella:2000bw}, 
and simulate the events by using the ATLFAST
detector simulator\cite{ATLFAST}. 
In
order to minimize the combinatorial backgrounds we use the event pairings
which satisfy the following conditions.
\begin{itemize}
\item Eq.~(\ref{maineq})
has solution for only one of the two possible lepton assignments.
\item
For the selected lepton assignment the resulting quartic equation in
$\mgl^2$ has only two solutions, and the difference of the gluino
masses for the two solutions is more than 100~GeV.  The smaller gluino
mass solution is chosen.
\end{itemize}
Note the selections  are rather 
phenomenological and they may introduce some bias to the reconstructed 
sparticle masses.

The $\mgl$ distributions for the OSSF$\times$OSSF events pairs are
shown in the histograms on the upper line of Fig.~\ref{mglplot}.  
We find  the obtained peak positions are
consistent with the 
input masses in this study.
A significant SUSY background, also shown in Fig.~\ref{mglplot} is still
present in the sample.  This background can be estimated from the data
themselves by using the $bb\ell\ell$ events with an opposite sign
opposite flavor (OSOF) lepton pair (i.e. $\ell\ell=e^{\pm}\mu^{\mp}$).

The histogram shows a peak
corresponding to the input value for the gluino mass even before the
background subtraction.  The green and blue histograms show the
estimated background 
distributions. 
The distributions after the background subtraction are shown in the
histogram on the lower line of Fig.~\ref{mglplot}.  The peak position
and its error obtained by a Gaussian fit to the distribution are
and it is 591.9$\pm$0.7~GeV at SPS1a. 
One can also look into the distribution of
$m_{\tilde{g}}-m_{\tilde{b}}$ and estimating the value of this
observable by performing a Gaussian fit on the observed peak and 
the peak position is 98.9~GeV, while the input is 103.3~GeV.

The error of the peak position is O(1)~GeV, but this is not the error
of the gluino mass, because each event can be used many times for the
mass calculation. To estimate the error, we generate many signal
events at SPS1a($\tan\beta=10$) and study the deviation of the peak
positions of the sub-samples. The gluino mass error for $\int dt {\cal
L} =300$fb$^{-1}$ is $\Delta m_{\tilde{g}}\sim 2$~GeV at SPS1a.

\begin{figure}[thb]
\centerline{
\includegraphics[width=5cm]{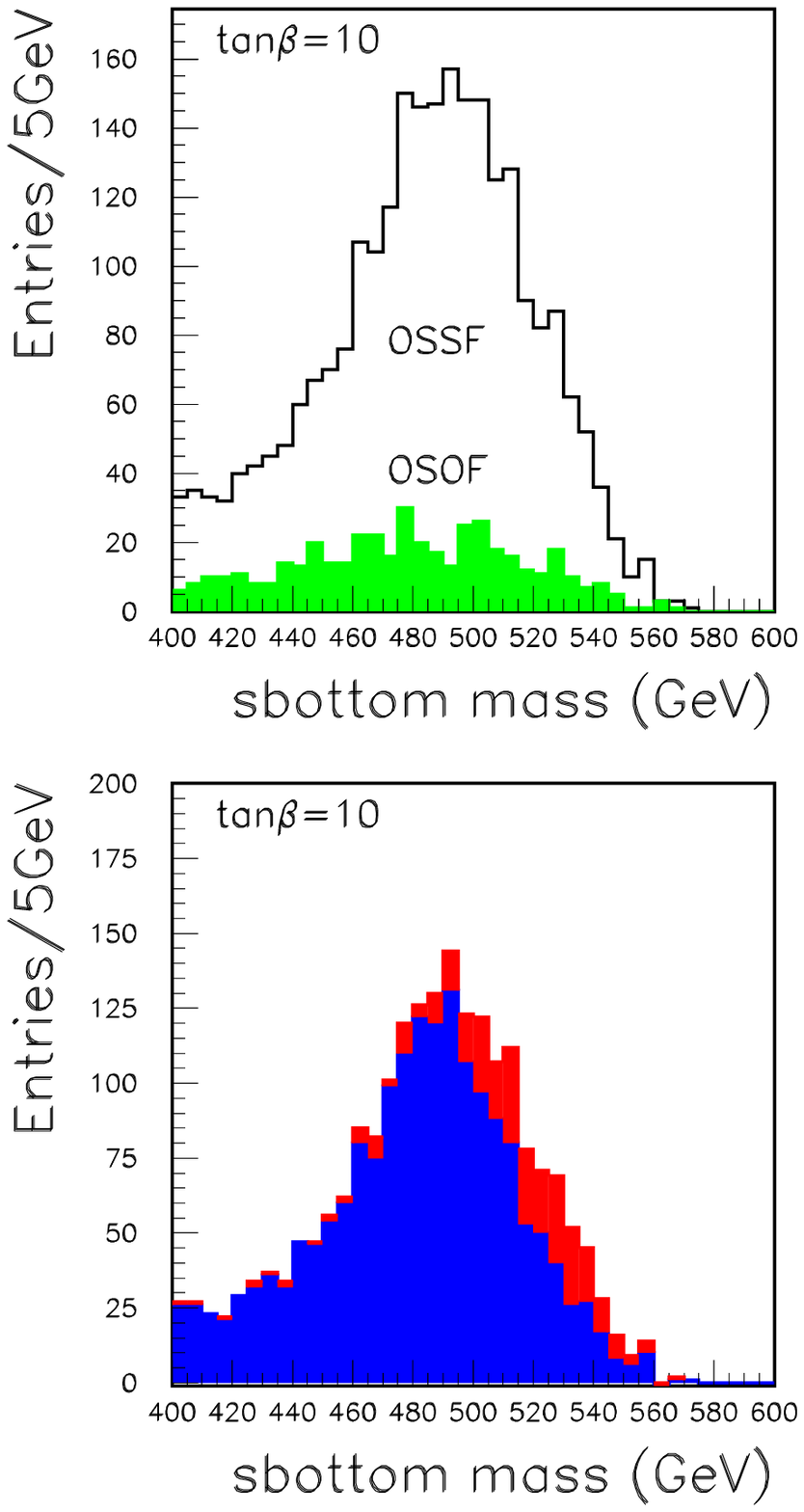}
\includegraphics[width=5cm]{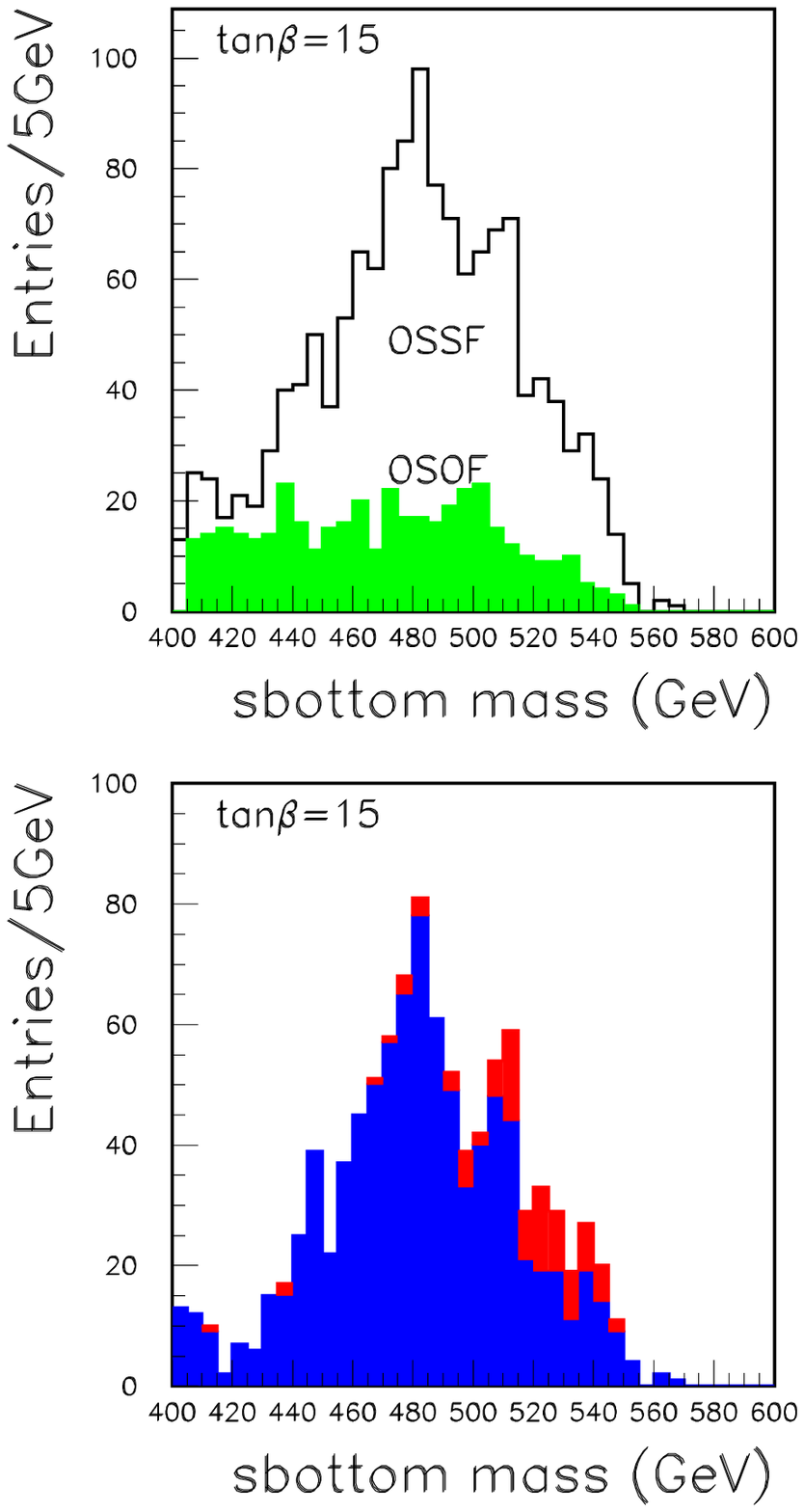}
\includegraphics[width=5cm]{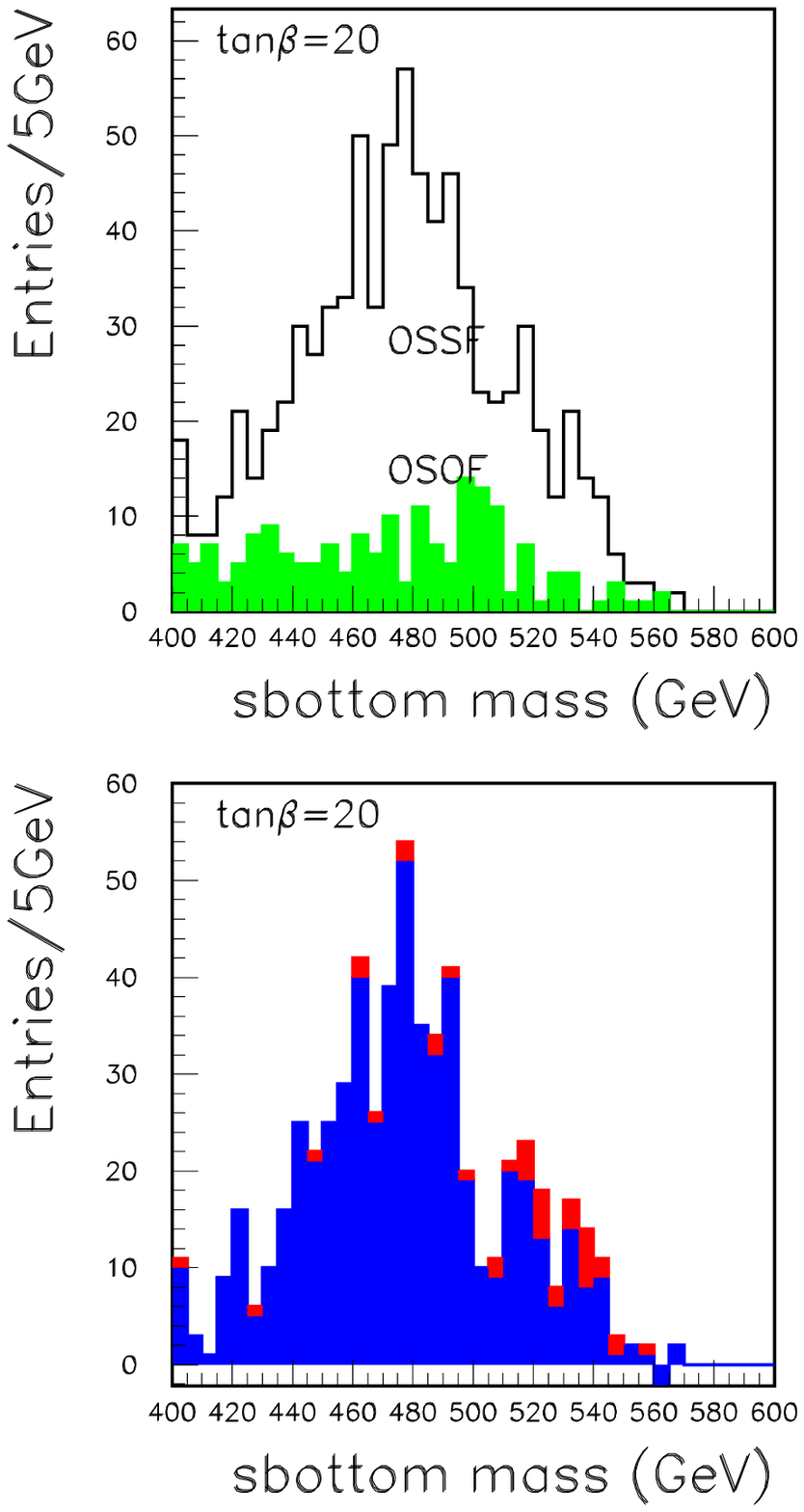}
}

\caption{The $m_{\tilde{b}}$ distributions
for $\tan\beta=10$ (left), 15 (center)  and 20 (right)
with a fixed gluino mass
($\mgl=595$~GeV).
The open and green histograms in the top figures
show the distributions of OSSF and OSOF lepton events, respectively.
The mass distributions after background subtraction
are shown in the bottom figures,
where the contribution of the $\tilde{b}_2$ events are shown
by the red regions.
}
\label{smallmsb}
\end{figure}

Once the gluino mass is fixed by the analysis shown above,
Eq.~(\ref{maineq}) can be solved for each event for 
two solutions for the sbottom mass,
giving as input the central measured value for the gluino mass.
In Fig.~ \ref{smallmsb}, we plot the
distribution of the smaller sbottom mass solution $m_{\tilde{b}}$(min)
for both OSSF (signal) and OSOF (background) lepton pair events (top
histograms) for SPS1a and two other model points which 
has same MSUGRA parameters but  $\tan\beta=15$ and 20.  
The relevant sparticles masses are listed in Table~\ref{table1}. 
The $m_{\tilde{b}_1}$ changes sensitively with $\tan\beta$ 
due to the increased left right mixing. 
We show in the bottom line the mass distributions after
the background subtraction. 
The peak positions are evaluated by a
Gaussian fit, and listed in Table~\ref{mglfit2}.  Note that the total
number of the signal $\tilde{b}_1$ event is smaller by factor of 4 for
$\tan\beta=20$ compared with $\tan\beta=10$. This is 
because the decay  $\tilde{\chi}^0_2\to \tau\tilde{\tau}$ dominates
as $\tan\beta$ increases. Thanks to the peak structure of the
distribution, the mass peak is still seen
very clearly.  The $m_{\tilde{b}}$(min) peak and
$m_{\tilde{b}}$(input) are in good agreement, and it is not so for the
larger solution $m_{\tilde{b}}$(max). 

The peak positions of the distribution of the events originated from
$\tilde{b}_2$ decay are also consistent with $\tilde{b}_2$ masses. 
The existence of $\tilde{b}_2$ can be established only after understanding
$b$ jet smearing and $\tilde{b}_1$ distribution correctly.

\begin{table}
\begin{center}
\begin{tabular}{|c||c|c|c|}
\hline
            &$\tan\beta=10$&   $\tan\beta=15$&   $\tan\beta=20$\cr
\hline
$\msb$(true)    &      491    &          485.3  &          478.8\cr
$\msb$(min)&  492.1$\pm$ 1.2& 487.7$\pm$ 2.2& 474.3$\pm$ 2.4\cr
$\msb$(max)&  504.5$\pm$ 1.0& 502.9$\pm$ 1.7& 495.1$\pm$ 2.4\cr
\hline
\end{tabular}
\caption{Fit results of the sbottom mass in GeV with a  fixed gluino mass
($\mgl=595$~GeV).
}
\label{mglfit2}
\end{center}
\end{table}

\subsection{Likelihood analysis}
Because  Eq.~(\ref{maineq}) have   multiple 
solutions for a fixed gluino mass, 
we encountered the problem to select the one of the 
multiple solution artificially in the previous 
subsection. To avoid bias to the analysis, 
we introduce a ``likelihood analysis'' in this subsection. 
\par From Eq.~(\ref{maineq}), each event is represented as 
a curve in the ($\mgl, \msb$) plane. The coefficients of the curve are 
a function of  the four momenta of the detected partons. 
The partons are measured as jets in the detector, which smears
the parton according to a smearing function. 
From the
measured quadratic form for an event we can build a confidence belt in
the ($\mgl, \msb$) plane\cite{Allanach:2004ub,Kawagoe:2004rz}.  
In order to build the ``probability
distribution function'', the crucial ingredient is the distribution of
the measured $b$-jet momenta as a function of the $b$-parton
momenta. 
We define an 
approximate probability density function according to the formula:
\begin{equation}
{\cal L}(\mgl,\msb)
=\int dp'_1 \int dp'_2 \epsilon(p_1:p'_1)\epsilon(p_2:p'_2)
\delta(f(\mgl,\msb,p'_1,p'_2))
\label{like}
\end{equation}
where $\epsilon(p:p')$ is the probability to measure a momentum $p'$ 
for a $b$ jet, given a $b$-parton with momentum $p$.

In Eq.~(\ref{like})  we did not include the possibility of 
lepton momentum mis-measurement, which has an almost negligible
effect. We also assume that the jet direction is not 
modified by the measurement and we use for $\epsilon(p:p')$ a gaussian 
distribution, with a width $\sigma$ 
corresponding to the parameterized jet smearing 
used in the fast simulation program. 
The approximate function takes however into account the
dominant part of the jet smearing and can be used to demonstrate the
method. 

\begin{figure}
\begin{center}
\includegraphics[width=4cm]{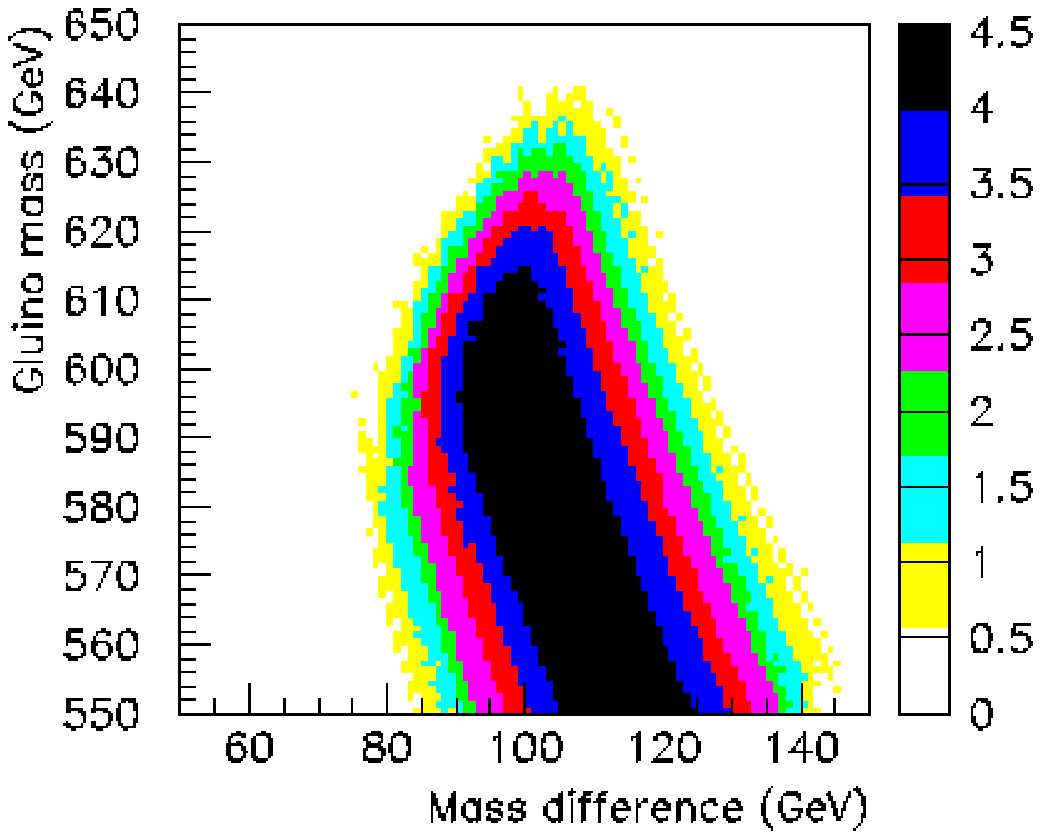}
\includegraphics[width=4cm]{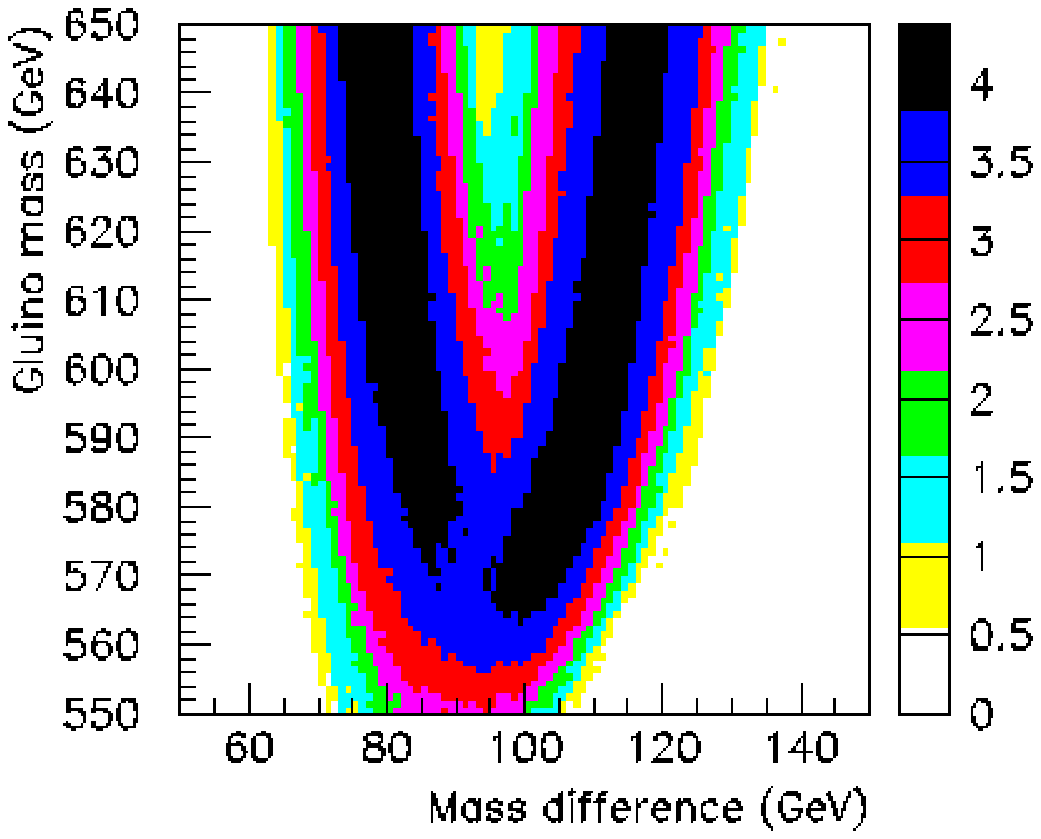}
\includegraphics[width=4cm]{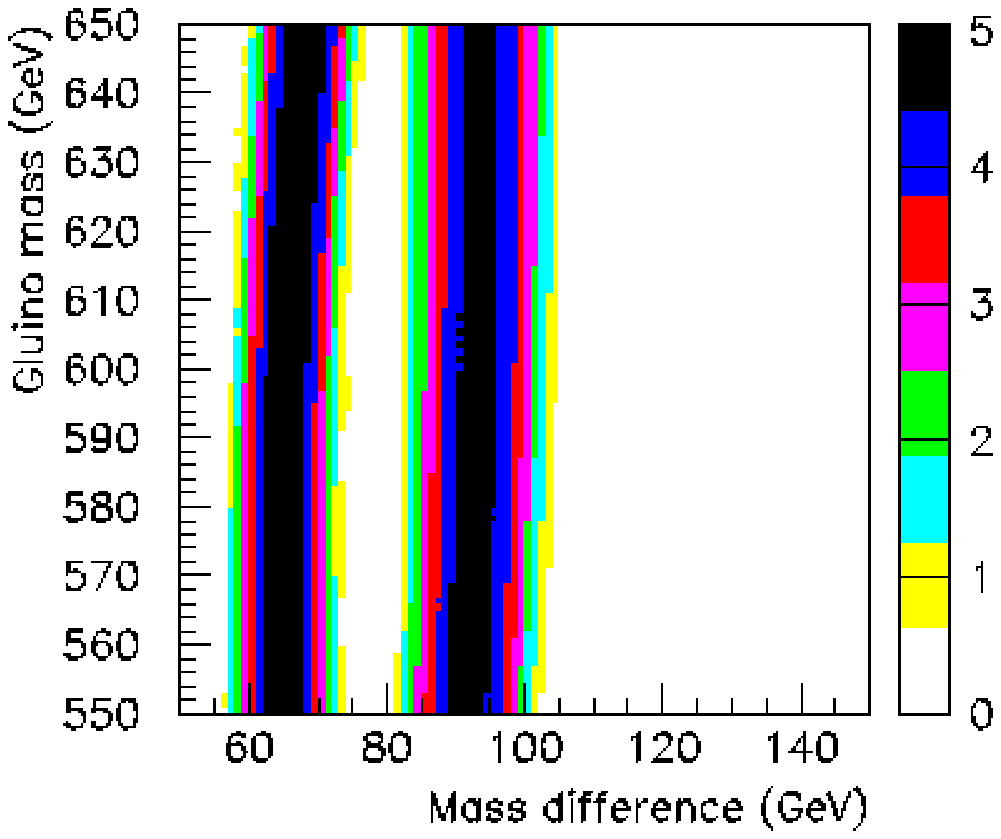}
\end{center}
\begin{center}
a)\hskip 4cm b) \hskip 4cm c) 
\end{center}
\caption{Likelihood distributions in the $(\mgl-\msb,\mgl)$
for selected events.}
\label{events}
\end{figure}

 We now show numerically calculated  $\log {\cal L}$ in the
$(\mgl-\msb,\mgl)$ plane for a few events where the $bb\ell\ell$
events originates from the cascade decay of Eq.~(\ref{bbll}) at
SPS1a. We calculate $\cal{L}$ using the following
procedure.  For each event in our
sample, characterized by a ($p_1,p_2$) pair of measured 
momenta for the $b$-jets, 
we generate Monte Carlo events of two $b$ jets with momentum 
($p'_1$, $p'_2$),                      
where $p'_1$ and $p'_2$ are
randomly generated according to the function $\epsilon(p_1:p'_1)
\times\epsilon(p_2:p'_2)$. 
The histgrams of the number of curves which satisfy Eq.~(\ref{maineq})
 that go through 
$1$~GeV$\times1$~GeV grid in the $(\mgl,\msb)$ plan, normalized by 
dividing the total generated event $n$ corresponds to 
${\cal L}(m_{\tilde{g}}, m_{\tilde{b}})$.
In Fig.~\ref{events},
we plot
\begin{equation}
\Delta \log {\cal L}
= \log( {\cal L}( \mgl, \mgl+\Delta\mgl, \msb,\msb+\Delta\msb)+ c)-
\log({\cal L}({\rm min})),
\label{evlik}
\end{equation}
where $c=0.001$ is a constant cutoff factor, which is 
needed as for each event we generate only 
a finite number of Monte Carlo experiments, 
and therefore some bins can have zero hits. 
The shape of the probability density distribution 
is different event by event, as it depends on the event kinematics.
The size of the  allowed band is also different,
which means that some events will have 
more weight in the determination of the mass parameters. 

By combining the probabilities for different events,
a region of maximum probability in the ($\mgl,\msb$) is found,
where the curves of maximum probability for all events 
approximately cross. 
Namely, we can build the combined likelihood for all the events 
defined as: 
\begin{equation}
\log {\cal L}_{\rm comb} ( \mgl, \mgl+\Delta\mgl,
\msb,\msb+\Delta\msb) = \sum_{\rm events} 
\log( {\cal L}( \mgl, \mgl+\Delta\mgl, \msb,\msb+\Delta\msb)+ c).
\end{equation}
To study the likelihood distribution of gluino and sbottom mass, 
we actually shows the distribution of subtracted likelihood
\begin{equation}
\log {\cal L}_{\rm sub}
\equiv \log {\cal L}_{\rm OSSF} - \log {\cal L}_{\rm OSOF}
\equiv
\sum_{\rm OSSF}\log{\cal L} 
-\sum_{\rm OSOF}\log{\cal L}.
\label{sub}
\end{equation}
This is not the correct definition of the likelihood function, but 
in the limit of infinite statistics, $\log {\cal L}_{\rm sub}$ 
is independent from the contribution of accidental lepton pairs.
For the correct treatment, see \cite{Kawagoe:2004rz}.

\begin{figure}
\includegraphics[width=8cm]{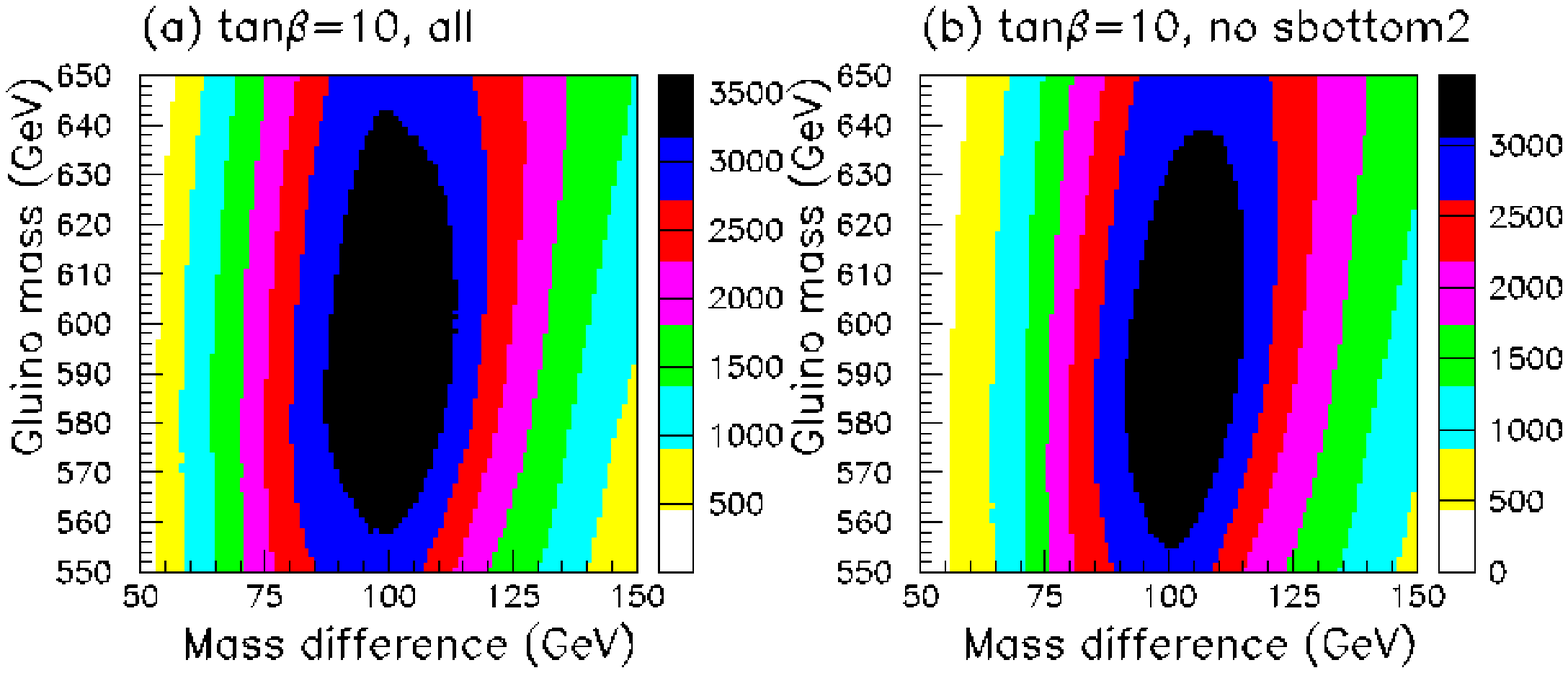}
\includegraphics*[width=8cm]{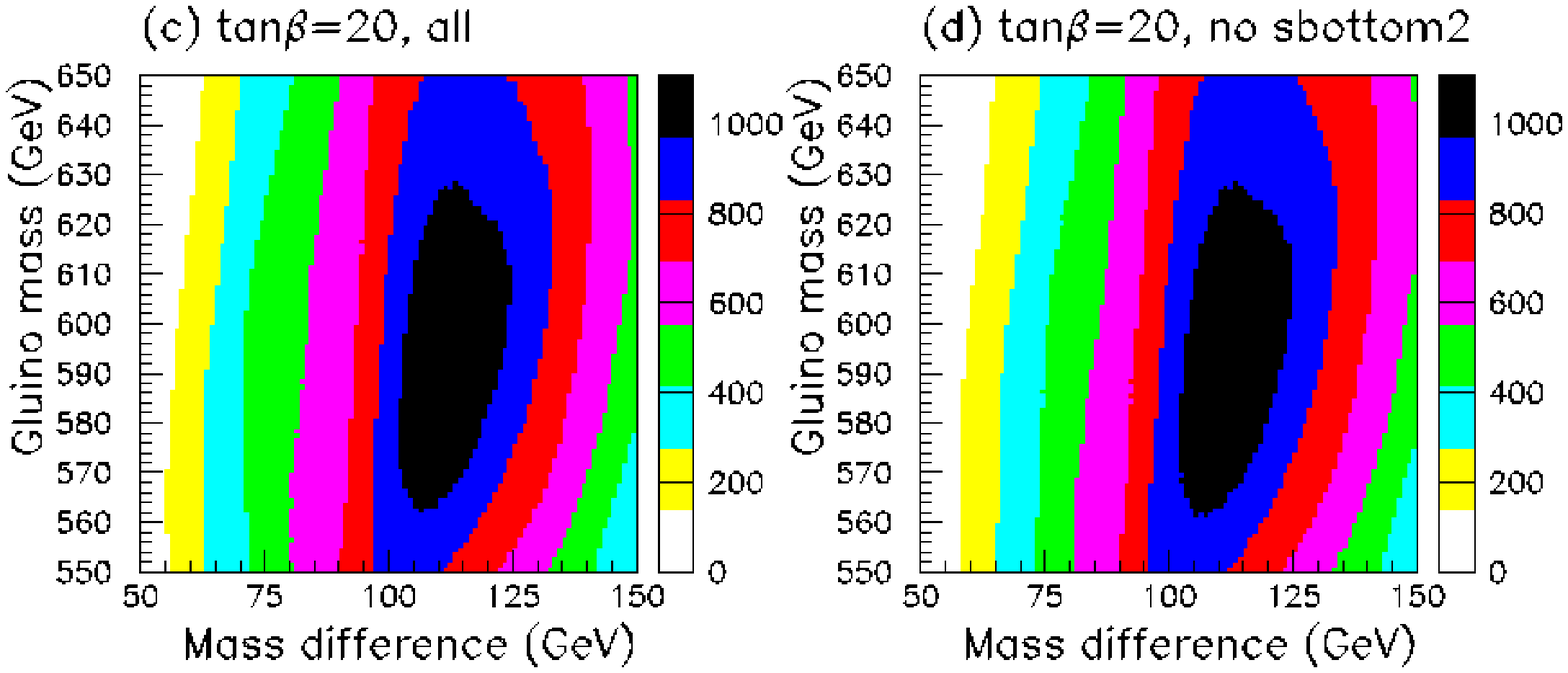}
\caption{Contours of the likelihood function
$\log {\cal L}_{\rm sub}$ in the $(\mgl-\msb,\mgl)$ plane:
(a)  for (c) $\tan\beta=10$ and (b) and (d) for $\tan\beta=20$,
respectively. 
The contours (b) and (d) are made without $\tilde{b}_2$ contributions.}
\label{nocut}
\end{figure}

We plot the contours of the function $\log {\cal L}_{\rm sub}$ in
Fig.~\ref{nocut}, where plots (a) and (b) [(c) and (d)] are for
$\tan\beta=10$ [$\tan\beta=20$].  The distributions (a) and (c) are
produced accepting all the events which pass the selections, whereas
distributions (b) and (d) are produced using an event sample where the
events including a $\tilde{b}_2$ decay have been rejected.

In Fig.~\ref{nocut} (a) and (c), the position of the peak for
$\mgl-\msb$ is roughly consistent with the input value.  Unlike the
gluino and sbottom mass fits in the previous section, we obtain the
correct peak position without the need of artificially choosing among
multiple solutions.  The likelihood distribution can be used to
determine the $\tilde{g}$ and $\tilde{b}$.  We restrict the likelihood
distribution for $591$~GeV $<m_{\tilde{g}}<599$~GeV (within 4~GeV from
the input gluino mass). We then fit the distribution around the peak
assuming gaussian distribution, The likelihood distribution peaks at
the gluino and sbottom mass difference as 99.5~GeV for $\tan\beta=10$,
104.2~GeV for $\tan\beta=15$, and 113.9~GeV for $\tan\beta=20$, where
the input value is 103.3~GeV, 109.9~GeV and 116.5~GeV, respectively.
The fitted values display shift of about 4~GeV from the true value. We
ascribe this effect to our simplified modeling of the jet smearing in
building the likelihood function, which should disappear once the
detector response is properly taken into account in the unfolding
procedure.

\begin{figure}
\begin{center}
\includegraphics[width=10cm]{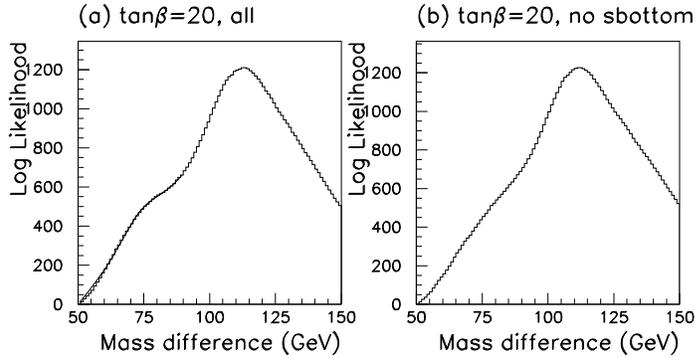}
\end{center}
\caption{The likelihood as a function of $\mgl-\msb$
for $\tan\beta=20$:
(a) for all events and
(b) without $\tilde{b}_2$ events.}
\label{tanb20}
\end{figure}

In Fig.~\ref{tanb20}(a), we show the distribution of $\log {\cal
L}_{\rm sub}$ as a function of $\mgl-\msb$ at $\tan\beta=20$,
restricting the gluino mass in the region $591$~GeV$< \mgl<$599~GeV
again.  On the left of the peak corresponding to the $\tilde{b}_1$
mass, we see a small bump in the distribution.  This bump is not
observed in the mass distribution made without $\tilde{b}_2$
contribution (Fig.~\ref{tanb20}(b)).  In order to claim the presence
of a second component in the distribution on the data, the ability of
correctly reproducing the likelihood distribution for $\tilde{b}_1$
events would be needed.  It is also difficult to extract a statistical
significance for the $\tilde{b}_2$ shoulder as our definition of the
likelihood function is approximate one. 

\subsection{More challenges}

In the previous subsection, we have discussed the MSSM parameter
determination at SPS1a. For this point, the mass spectrums would be
obtained quite accurately by looking into the decay cascades involving
leptons. The SUSY scale is relatively light $m_{\tilde{q}}\sim
m_{\tilde{q}}\sim 600$~GeV, when production cross section is as large
as 50~pb. One can even
determine the edges of the $\tilde{\chi}^0_4$ decays.  
If the squark and gluino mass scale is around 1 TeV, the
production cross section is only around 3~pb, and statistical errors
would be increased accordingly.

The sensitivity to the masses would be further reduced when
$m_{\tilde{q}}>m_{\tilde{g}}$. For this case squarks decay dominantly
into gluino, and the gluino decay is 3 body.  Then it is hard to
identify the jets from squark and gluino decays.  This is typically
the case for the model points in the ``Focus point region'' in the
supergravity model.

The mass resolution also may be  worse when $\tan\beta$ is 
large in MSUGRA,  because the branching ratios of the $\tilde{\chi}^0_2$ 
into $\mu$ or $e$  are reduced significantly as 
$Br(\tilde{\chi}^0_2\rightarrow\tilde{\tau}\tau)$ dominates. 
The decay into $\tau$ may also be studied, but 
a $\tau$ jet resolution is significantly worse compared to 
that of a lepton, and the background are much higher 
due to fake $\tau$'s from  QCD jets.
The case where $\tilde{\tau}$ and $\tilde{\chi}^0_1$ are degenerated
in mass  
is favored in MSUGRA model with the dark matter  
constraint, because the coannihilation of $\tilde{\tau}$ and 
$\tilde{\chi}^0_1$ reduces the thermal relic density in the 
Universe to the acceptable level. In that case, one of the $\tau$ lepton 
from $\tilde{\chi}^0_2$ decays may  be too soft to be 
detected at the LHC. 

\begin{figure}
\begin{center}
\includegraphics[width=5cm]{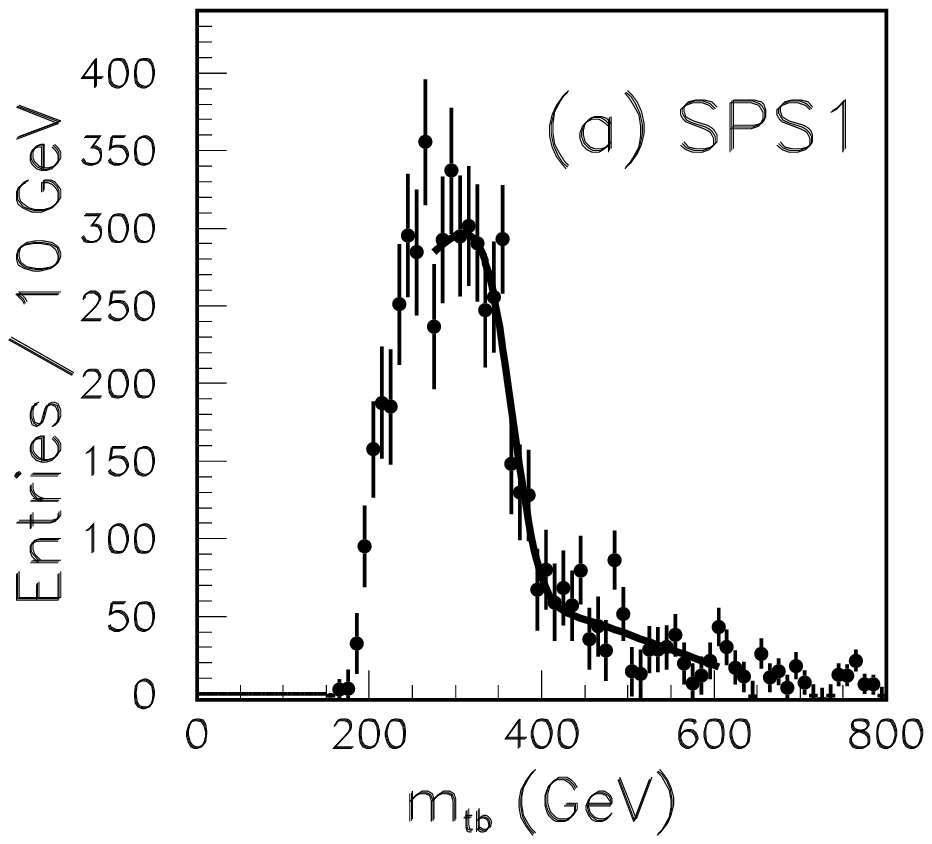}
\includegraphics[width=5cm]{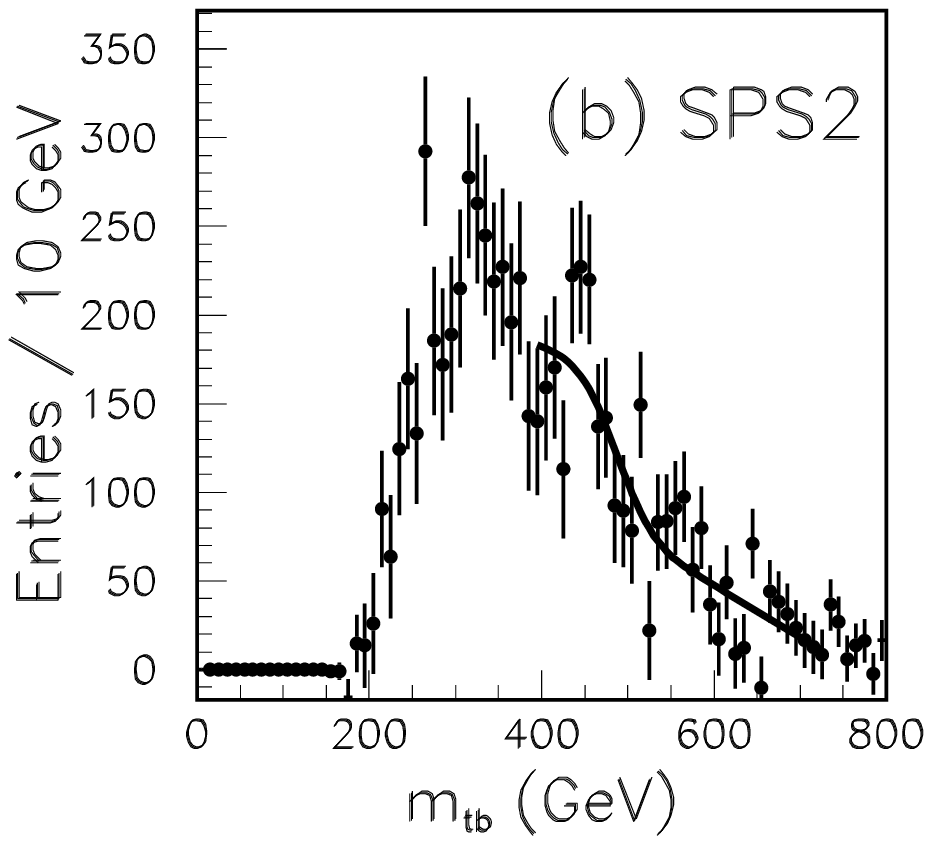}
\end{center}
\caption{The $m_{tb}$ distribution at SPS1a and SPS2. 
The formar is dominated by the two body cascade decay 
$\tilde{g}\to t\tilde{t}, b\tilde{b}\rightarrow tb\tilde{\chi}^+_1$ 
and the latter is dominated by the three body decay 
$\tilde{g}\to tb\tilde{\chi}^+_i$. }
\label{mtb}
\end{figure}

It is therefore important to establish analysis which does not relay
on the leptons in the event cascades. For this purpose we discussed
the reconstruction of the $tb$ final state where top decays
hadronically\cite{Hisano:2003qu}.  The dominant decay of gluinos for
these points are $\tilde{g}\rightarrow
(\tilde{t},\tilde{b})$$\rightarrow t b \tilde{\chi}^\pm_i$. The top
quarks in the SUSY events may be reconstructed by looking for the
$bjj$ where $m_{jj}\sim m_W$ and $m_{bjj}\sim m_t$. There are many
accidental background which satisfy this conditions, because the
number of jets $n_{\rm jets}$ in the SUSY events are typically 8 to
10. Such backgrounds can be subtructed by using the side band events
where $m_{jj}<m_W-15$~GeV and $m_{jj}>m_W+15$~GeV. The reconstructed
top quarks are then used to study the $m_{tb}$ distributions. In
Fig.~\ref{mtb}, we show the $m_{tb}$ distribution after the background
subtruction for SPS1a and SPS2, where SPS2 corresponds to the focus
point where $m_{\tilde{q}}\sim 800$~GeV and
$m_{\tilde{q}_L}\sim1.5$~TeV. For SPS1a, $\tilde{g}\rightarrow
\tilde{t}t$ and $\tilde{b}b$ is open, and distribution shows 
sharp edge consistent to those calculated from the input
masses. The endpoint is measured within an
error $\Delta m_{tb}\sim 4$~GeV, and the distribution maybe used to
study the $\tilde{t}$ nature.  For the case of SPS2, the decay chain
is not open, but gluino still dominantly decay into third generation
particles because stop and sbottom masses are significantly lighter
than that of the first and second generation squarks. The distribution
does not show the edge strucuter because the gluino decay is  three
body.  The $m_{tb}$ endpoint is consistent to the mass difference
$m_{\tilde{g}}-m_{\tilde{\chi}^+_1}\sim 560$~GeV and
$m_{\tilde{g}}-m_{\tilde{\chi}^+_2}\sim 480$~GeV.

\section{Supersymmetric Relations}

In MSSM, all dimensionless 
couplings involving supersymmetric particles 
are described by either SM Yukawa  or  gauge couplings. 
Not only the the size of the couplings, but 
the chiral structure of the couplings 
are restricted by the supersymmetry.  
The fermion-sfermion-gaugino coupling 
is restricted to preseve the chirality of (s)fermion,  therefore 
for example, 
\begin{equation}
{\cal L}=\frac{g}{\sqrt{2}}\left[\overline{\tilde{W}} l_L \tilde{l}^{*}_L
+\tan\theta_W\overline{\tilde{B}} l_L \tilde{l}^{*}_L
-2 \tan\theta_W\overline{\tilde{B}} f_R\tilde{f}^{*}_R + h.c.\right],
\end{equation}
for the lepton-slepton-gaugino interaction, while the chirality 
must be flipped for the Higgsino interactions. 
Proving the   couplings and their chiral structures 
is important steps to 
check if the  discovered new particles are indeed superpartners
of the SM particles or not.  

For some cases one can check the chiral structure of 
the interaction by looking into the decay distributions. 
We first note that  the SUSY particles 
are naturally  polarized in many SUSY process. 
For example, in the cascade decay 
\begin{equation}
\tilde{q}_L\rightarrow \tilde{\chi}^0_2q_L\rightarrow \tilde{l}_Rl_Rq 
\to llq\tilde{\chi}^0_1,
\end{equation}
the $\tilde{\chi}^0_2$ is 
polarized as right-handed (opposite to $q_L$), because 
the Yukawa coupling of squark-quark-ino flip the chirality. 
The polarized $\tilde{\chi}^0_2$
further decay into  either leptons-antislepton,
or antilepton-slepton. The two decay branching ratios are same 
because $\tilde{\chi}^0_2$ is a Majorana 
particles. ¡¡For the decay  $\tilde{\chi}^0_2\rightarrow
\tilde{l}_Rl^+$
through Yukawa type interaction, 
anti-lepton likely to  go in the same direction 
to $\tilde{\chi}^0_2$ (namely opposite to the jets 
in the squark rest flame), while 
the  lepton goes in the same direction to the jet 
for the  $\tilde{\chi}^0_2\rightarrow \tilde{l}_R^* l^-$ decay. 

The difference of the angular distribution appears 
as the charge asymmetry in the  $m(jl)$
distribution\cite{Richardson:2001df}, becuase the  $m(jl)$ is propotional to $1-\cos\theta$,
where $\theta$ is the angle between the jets and the lepton 
in the $\tilde{\chi}^0_2$ rest flame. The 
$m(jl^+)$ is peaked sharply at the endpoint of  $m(jl)$ distirubtion, 
while the  $m(jl^-)$ distribution is suppresed at the endpoint. 
Finally  a (anti)lepton is 
emitted from a (anti)slepton, but this time 
the lepton angle distriution 
is spherical in the $\tilde{l}$ rest flame.

The charge asymmetry is exactly opposite for $\tilde{q}^*_L$
decays. However, at  $pp$ colliders, the number of produced squark is
much larger than that of anti-squark, and the $\tilde{\chi}^0_2$ produced
from $\tilde{q}_L$ decay dominates total $\tilde{\chi}^0_2$ productions.
The charge asymmetry in $m(jl)$ distribution in the $jll$ sample
therefore remains. From the distribution, one can conclude that the
decay products $\tilde{\chi}^0_2$ is a fermionic particle, and
the interaction is chiral\cite{Barr:2004ze,Goto:2004cp}
.

\begin{figure}[tbh]
\begin{center}
\includegraphics[height=4.5cm,clip]{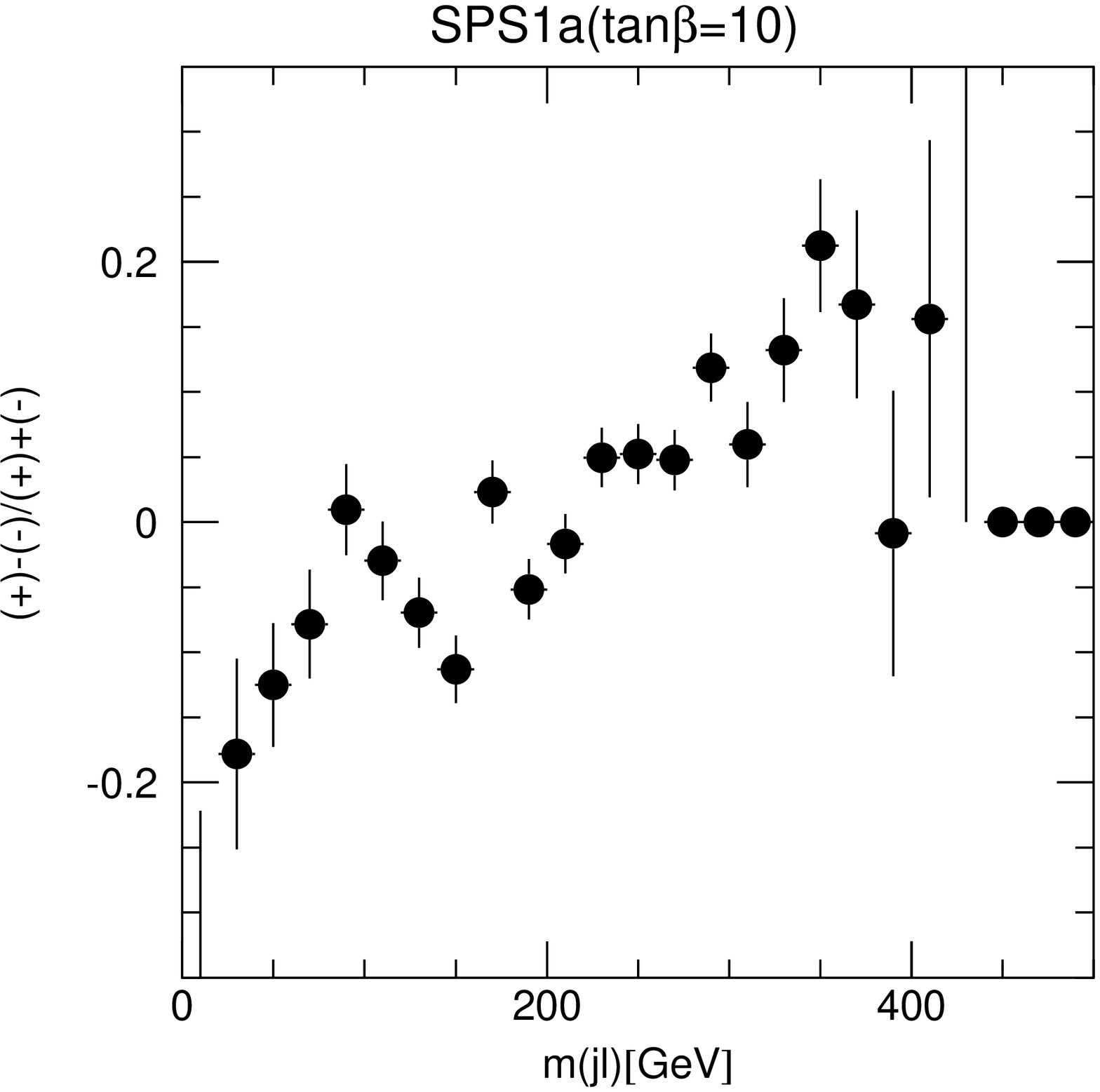}
\includegraphics[height=4.5cm,clip]{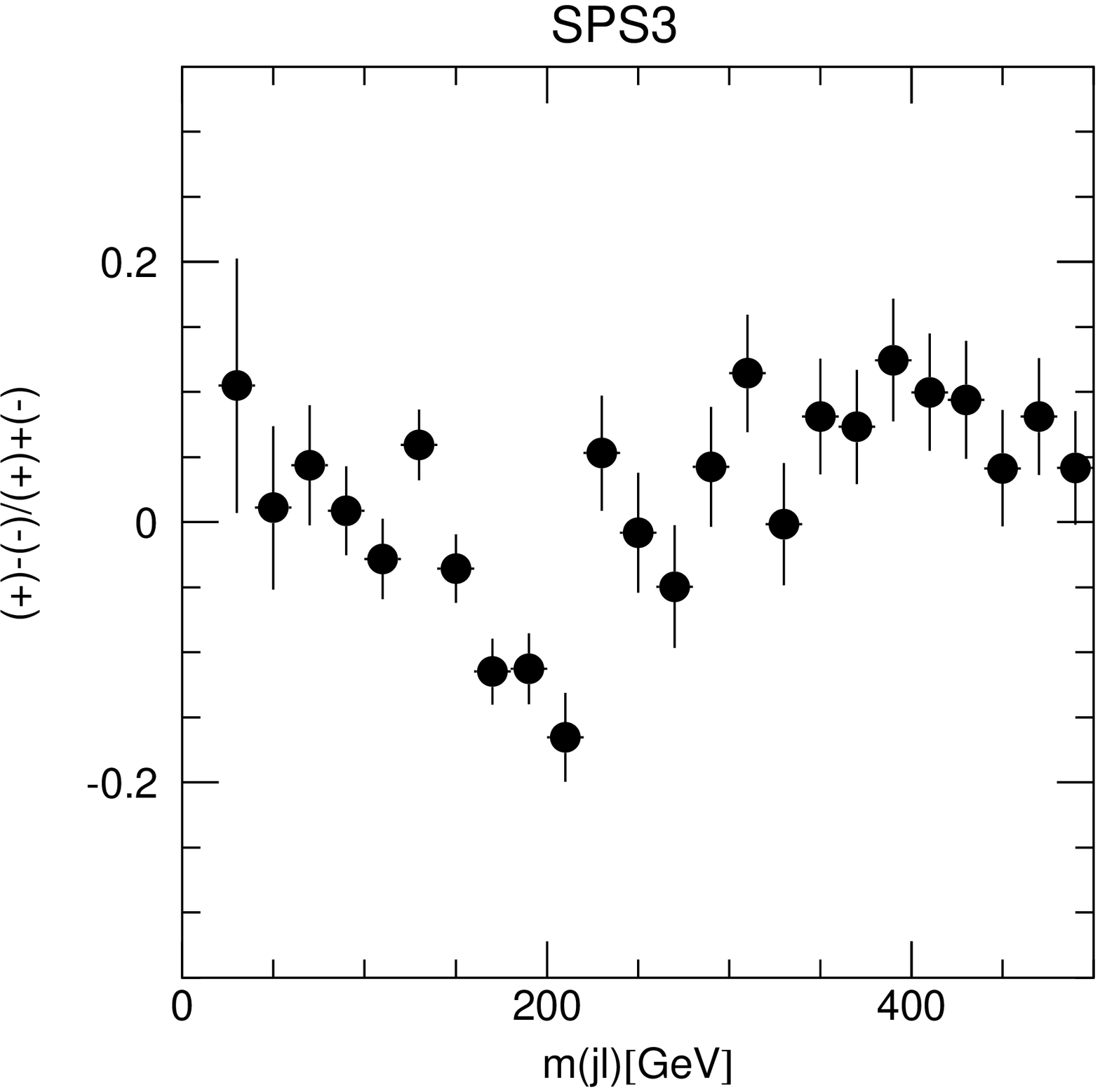}
\\
\end{center}
\caption{
 The reconstructed asymmetry 
$[N_{\rm sig}(jl^+)-N_{\rm sig}(jl^-)]
/[N_{\rm sig}(jl^+)+N_{\rm sig}(jl^-)]$.
}
\label{asymmetry}
\end{figure}

The detectability of the charge asymmetry can be studied by generating
events by using HERWIG\cite{Corcella:2000bw}, 
and simulate the events by using the ATLFAST
detector simulator\cite{ATLFAST}. 
Here we show two representative case at the points SPS1a and SPS3
in Fig.~\ref{asymmetry}, where the decay 
$\tilde{\chi}^0_2\rightarrow l\tilde{l}_R$
 dominates at SPS1a and $l\tilde{l}_L$ dominates at SPS3 respectively.
The distributions shows opposite asymmetry near the 
endpoints the $m(jl)$ distributions for the decay $\tilde{q}\rightarrow
q\tilde{\chi}^0_2 \rightarrow ql\tilde{l}$
(298~GeV at SPS1a and 200~GeV at SPS3 respectively).
This is because the chirality of the $\tilde{\chi}^0_2$-$l$-$\tilde{l}$ 
coupling is exactly opposite for the two 
points\cite{Goto:2004cp}. 

The cascade decay may also be  sensitive to the 
$\tilde{l}_L$ and $\tilde{l}_R$ mixing\cite{Goto:2004cp}. Due to 
the $F$ term contribution of slepton mass matrix, 
$m^2_{LR}\sim m_f\mu\tan\beta$, the lighter slepton 
mass eigenstates $\tilde{l}_1$ is
mixture of $\tilde{l}_L$ and $\tilde{l}_R$ 
where the mixing angle is defined as 
$\tilde{l}_1=\tilde{l}_L\cos\theta_l +\tilde{l}_R\sin\theta_l
$. 
The left and right hand coupling of the $\tilde{\chi}^0_2$ 
to the lighter  mass eigenstate  $l_1$
is expressed as 
\begin{eqnarray}
{\cal L}&=&
 -\frac{g_2}{\sqrt{2}}\tilde{\chi}^0_2 (C_L P_Ll+ C_R P_Rl)\tilde{l}^*_1,\cr 
C_L&\equiv&-\cos\theta_{l} (N^*_{\tilde{W}2}+ N^*_{\tilde{B}2}\tan\theta_W),
\cr
C_R&\equiv&2\tan\theta_W\sin\theta_{l}N_{\tilde{B}2},
\end{eqnarray}
when $\tilde{\chi}^0_2$ is gaugino like.

When $\tilde{\chi}^0_2\sim\tilde{W}$, $\tilde{\chi}^0_1\sim$
$\tilde{B}$, and $\tilde{l}_1\sim \tilde{l}_R$, $C_L$ is suppressed by
small mixing angle proportional to $m_f\mu\tan\beta$, while $C_R$ is
suppressed by the small bino component of the wino like
$\tilde{\chi}^0_2$. The non-universality appears as the sign of the
non-zero $F$ term mixing of $\tilde{\mu}$ and visible in the wide
parameter region.

While the mixing angle of $\tilde{\mu}$ is only around 1\% for the Snowmass 
point SPS1a, the difference of the decay branching ratio 
$Br(\tilde{\chi}^0_2\rightarrow\tilde{\mu}\mu$)$/Br(\tilde{\chi}^0_2\rightarrow
\tilde{e}\mu)$ from 1 is as large as  4\%. Although the 
statistics reduces as $\tan\beta$ increases, statistical significance 
of the signal defined as 
$$
S=\left(\frac{Br(\mu)}{Br(e)}-1\right)
\left(\frac{\Delta N({\rm sig})}{N({\rm sig})}\right)^{-1}
$$ 
where $\Delta N$(sig)$=\sqrt{N_{\rm OS}}$, and $N_{\rm OS}$
is the number of the total odd sign two lepton events. We found that 
$S$
increase from  5.4 at SPS1a to  8.5 for a modified point with 
$\tan\beta=20$ (see Table~\ref{sigbg}). 

\begin{table}
\begin{center}
\begin{tabular}{|c||c|c|c|c|c|}
\hline
point & $Br(\tilde{\chi}^0_2\to e\tilde{e})$
 & $N_{\rm sig}$($e$ and $\mu$)  & $N_{\rm OS}$& 
$\left(\frac{Br(\mu)}{Br(e)}-1\right)$& $S$  
\\
\hline
$\tan\beta=10$ & 6.3\% & $1.39 \times 10^4$ & \ $2.68 \times 10^4$\ &4\%& 5.4 \\
\hline
$\tan\beta=20$ & 1.2\%& $0.28 \times 10^4$ & \ $1.02\times 10^4$\ &17\%&8.5   \\
\hline
\end{tabular}
\end{center}
\caption{
The $Br(\tilde{\chi}^0_2\to e\tilde{e})$ and accepted number of events
for SPS1a $\tan\beta=10,20$.
$N_{\rm sig}$ is the number of $l^+l^-$ events after $e^{\pm}\mu^{\mp}$
subtraction, while $N_{\rm OS}$ is the number of 
the total odd sign two lepton events.
The expected deviation of $Br(\tilde{\chi}^0_2\rightarrow \tilde{\mu}\mu)$
from that for $\tilde{e}e$ is compared with the statistical error of the number of  the 
$e^+e^-$ events for $\int dt{\cal L}=300$ fb$^{-1}$.}
\label{sigbg}
\end{table} 

\section{Discussion}
In this talk, we discussed SUSY studies at the LHC. In
early '90, the LHC was considered  merely as a discovery machine,
due to the challenging experimental environment.
However, it was recognized that 
the parameters of the MSSM can be determined at the LHC. 
Since then, many techniques have been developed to extract the
mass and coupling information.

When the gravitino is the LSP and the NLSP is long lived, the LHC works as
the machine for precision studies. This is because SUSY events
contains many leptons and photons whose momentum can be 
easily combined to solve sparticle masses.  For the case of the charged NLSP, 
the 
semi-stable charged particles would  be observed as a highly ionizing
charged tracks which goes through the detector.  The decay position
and decay time of the NLSP can be measured at the LHC, and the
gravitino interaction to the matter may be explored.  The cosmology
related to the gravitino LSP and the collider phenomenology 
have received lots of attentions.

When the LSP is the lightest neutralino, the sparticle 
mass determination using the
endpoint analysis is known to be very powerful.
For the most favorable case, one can determine the LSP mass
within 5\% and squark and gluino masses within 2\%. 
We have also introduced  the mass relation method developed very 
recently in this talk. By
using this method, one can reconstruct sparticle masses as peaks in
the distributions calculated from the events using the exact
formula to solve decay kinematics. The mass relation method 
may be useful even if the event statistics is small.

Finally we discussed the study of the 
sparticle interaction at the LHC. The chiral structure of the 
sparticle couplings may be studied through  the  charge asymmetry of 
the selected $m(jl)$ 
distributions, 
where the   jet and the lepton come from the cascade decay 
$\tilde{q}_L\rightarrow \tilde{\chi}^0_2 q \rightarrow \tilde{l}lq$. 

To complete the task to explore all  masses and couplings of SUSY 
particles  is probably impossible by the LHC alone.
However even after the ILC is started, the 
measurements at the LHC are useful to understand  supersymmetry
in nature. There are many on-going studies, and 
we still need more ideas and thoughts
so that we do not miss any new physics signatures at the LHC.

\end{document}